\documentclass[11pt]{article}

\usepackage{amssymb}
\usepackage{mathrsfs}
\usepackage{microtype}
\usepackage{chet}

\newcommand{\DeltaW}{\Delta^{\hspace{-2pt}\text{W}}}
\newcommand{\dtwo}{\ensuremath{d^{\hspace{0.5pt}2}\hspace{-0.5pt}}}
\newcommand{\dfour}{\ensuremath{d^{\hspace{0.5pt}4}\hspace{0.5pt}}}
\newcommand{\dsix}{\ensuremath{d^{\hspace{1pt}6}}}
\newcommand{\dd}{\ensuremath{d^{\hspace{1pt}d}}}
\newcommand{\st}{\mathscr{T}}
\newcommand{\sz}{\mathscr{Z}}
\newcommand{\cA}{\mathcal{A}}
\newcommand{\cB}{\mathcal{B}}
\newcommand{\cC}{\mathcal{C}}
\newcommand{\cD}{\mathcal{D}}
\newcommand{\cE}{\mathcal{E}}
\newcommand{\cF}{\mathcal{F}}
\newcommand{\cG}{\mathcal{G}}
\newcommand{\cH}{\mathcal{H}}
\newcommand{\cI}{\mathcal{I}}
\newcommand{\cO}{\mathcal{O}}

\SetFootnoteHook{\hangindent=-5pt\noindent}
\DeclareNewFootnote[para]{EE}[roman]
\renewcommand{\emails}[1]{\footnotetextEE{\texttt{#1}}}

\date{August 2013}

\preprint{UCSD-PTH-13-12}

\title{Consequences of Weyl Consistency Conditions}

\author{Benjam\'in Grinstein, Andreas Stergiou and David Stone
\emails{\href{mailto:bgrinstein@ucsd.edu}{bgrinstein@ucsd.edu},
(\href{mailto:stergiou@physics.ucsd.edu}{stergiou},
\href{mailto:dcstone@physics.ucsd.edu}{dcstone})@physics.ucsd.edu}}

\affiliation{Department of Physics, University of California, San
Diego, La Jolla, CA 92093, USA}

\abstract{The running of quantum field theories can be studied in detail
with the use of a local renormalization group equation.  The usual
beta-function effects are easy to include, but by introducing
spacetime-dependence of the various parameters of the theory one can
efficiently incorporate renormalization effects of composite operators as
well.  An illustration of the power of these methods was presented by
Osborn in the early 90s, who used consistency conditions following from the
Abelian nature of the Weyl group to rederive Zamolodchikov's $c$-theorem in
$d=2$ spacetime dimensions, and also to obtain a perturbative $a$-theorem
in $d=4$. In this work we present an extension of Osborn's work to $d=6$
and to general even $d$.  We compute the full set of Weyl consistency
conditions, and we discover among them a candidate for an $a$-theorem in
$d=6$, similar to the $d=2,4$ cases studied by Osborn.  Additionally, we
show that in any even spacetime dimension one finds a consistency condition
that may serve as a generalization of the $c$-theorem, and that the
associated candidate $c$-function involves the coefficient of the Euler
term in the trace anomaly.  Such a generalization hinges on proving the
positivity of a certain ``metric'' in the space of couplings.}

\begin{document}

\maketitle

\newsec{Introduction}
When a symmetry of a quantum field theory (QFT) is broken by quantum
corrections, then the corresponding anomaly can be reproduced by a
contribution to the generating functional of the theory~\cite{Wess:1971yu,
Witten:1983tw}.  The algebra of the symmetry that is violated constrains
the symmetry-breaking parameters that appear in these anomalous
contributions, which are thus forced to satisfy the so-called Wess--Zumino
consistency conditions~\cite{Wess:1971yu}.

The study of the Wess--Zumino consistency conditions for the Weyl anomaly
was undertaken by Osborn in the early 90s and produced remarkable
results~\cite{Osborn:1991gm}.  In $d=2$ spacetime dimensions, for example,
Osborn obtained an independent proof of Zamolodchikov's
$c$-theorem~\cite{Zamolodchikov:1986gt}.  Furthermore, an extension of the
$c$-theorem to 4d, commonly referred to as the $a$-theorem, was
demonstrated perturbatively~\cite{Jack:1990eb, Osborn:1991gm}, establishing
in perturbation theory the intuition that the number of massless degrees of
freedom of a QFT decreases under renormalization-group (RG) flow.

This perturbative 4d result was based on previous work by Jack and
Osborn~\cite{Jack:1990eb}, who computed the local RG equation for a general
renormalizable QFT in a curved background using dimensional regularization.
To account for effects of renormalization of composite operators, Jack and
Osborn used spacetime-dependent coupling constants, a trick that allows for
straightforward computations of Green functions of composite operators (at
least of those that appear in the Lagrangian) and the stress-energy tensor.
Their candidate $c$-function agrees with Cardy's
suggestion~\cite{Cardy:1988cwa}: it is equal to the coefficient $a$ of the
Euler term in the trace anomaly at fixed points of the RG-flows.
Nevertheless, it differs from $a$ if the corresponding flat-space theory is
not a conformal field theory (CFT). In subsequent work Osborn reproduced
the main results of~\cite{Jack:1990eb} by requiring that two successive
Weyl variations of the effective action commute (since the Weyl group is
Abelian), and also showed that the main results of the analysis are
scheme-independent~\cite{Osborn:1991gm}.

Invariance under the Weyl group in the flat background limit is {\it a
priori} a stronger requirement than that of scale invariance of a theory on
a flat background. The former imposes the vanishing of the trace of the
stress-energy tensor, while the latter requires that the trace be the
divergence of a suitable vector operator~\cite{Polchinski:1987dy}. This
suggests that the study of Weyl deformations may elucidate the relation
between scale and conformally invariant theories in flat backgrounds.
Indeed, these methods have recently been used to show, in perturbation
theory, that a unitary 4d QFT invariant under the Poincar\'e group extended
by the generator of scale transformations is automatically invariant under
the four generators of special conformal
transformations~\cite{Fortin:2012hn}, even though the parameters of the
theory may display cyclic behavior~\cite{Fortin:2012cq}.\foot{The same
result was also obtained using different methods in~\cite{Luty:2012ww}. In
renormalizable theories with $\mathcal{N}=1$ supersymmetry it was shown
perturbatively that cyclic behavior of parameters does not
arise~\cite{Fortin:2012hc, Nakayama:2012nd}. The situation in
$d=4-\epsilon$ needs further investigation~\cite{Fortin:2011ks,
Fortin:2011sz, Fortin:2012ic}.}

In this paper we undertake an investigation of the response of QFTs to Weyl
transformations in $d=6$.\foot{Weyl consistency conditions in $d=3$ were
studied recently in \cite{Nakayama:2013wda}.} In particular, we determine
the Weyl consistency conditions and general RG properties of
six-dimensional QFTs. It is evident that this line of inquiry is
interesting if it leads to results similar to those already obtained in
$d=4$, say, a perturbative extension of the 2d $c$-theorem and a proof that
scale implies conformal invariance in 6d.  But the investigation is also
interesting in light of the advent of strongly coupled conformal CFTs that
lack a Lagrangian description in $d=6$, like the famous $(2,0)$
theory.\foot{The consistency conditions we derive can be seen as relations
among correlation functions involving composite operators.}  This suggests
the existence of flows, i.e.\ families of non-conformal QFTs, between such
theories, with an associated flow of a presumed $c$-function.  The class of
perturbative, renormalizable, $d=6$ models is restricted to scalar fields
with cubic interactions for which the general analysis is of limited
interest. They could, however, be put to use as a check of results using
standard methods of perturbation theory.

As we will see, the consistency conditions bring us very close to an
extension of the $c$-theorem to $d=6$. More specifically, one can define a
quantity $\tilde a$, which is a function of the dimensionless coupling
constants $g^i$, that satisfies the equation
\eqn{\frac{d\tilde{a}}{dt}=-\cH_{ij}\beta^i\beta^j,}[atheorem]
where the RG time is $t=-\ln(\mu/\mu_0)$, taken here to increase as we flow
to the IR, and $\beta^i=-dg^i/dt$, as usual.  The quantity $\tilde a$
agrees with the coefficient of the Euler term in the 6d trace anomaly at
fixed points.  The symmetric tensor $\cH_{ij}$ can be viewed as a
metric in the space of couplings. A proof of the $a$-theorem would be
immediate if $\cH_{ij}$ were shown to be positive-definite. This is
analogous to the situation in $d=4$ where perturbative positivity of the
analogous ``metric'' has been shown by explicit computation in a generic
QFT---it is here and only here that perturbation theory is used in proving
the $a$-theorem and that scale implies conformal invariance in $d=4$.

The analysis of the Weyl consistency conditions in $d=6$ is significantly
more complicated than in $d=4$. This analysis reveals generic features that
were not apparent in Osborn's treatment, and actually allows us to
demonstrate the validity of \atheorem for QFTs in any even-dimensional
spacetime.  The only ingredient missing for a generalization of
Zamolodchikov's $c$-theorem to any even dimension is a demonstration that
the ``metric'' $\cH_{ij}$ is positive-definite.  As already mentioned,
$\cH_{ij}$ is positive-definite in $d=4$ at lowest order in perturbation
theory. It would be interesting to extend this result to higher even
dimensions. Of course, a non-perturbative proof of the positivity of
$\cH_{ij}$ in even $d\ge4$ is the ultimate goal of this line of research.

To address questions similar to those that motivated this work, Komargodski
and Schwimmer have put forward an argument that gives a non-perturbative
physicist's proof of the weak version\foot{For a discussion of the various
versions of the $a$-theorem see \cite{Barnes:2004jj}.} of the $a$-theorem
in $d=4$~\cite{Komargodski:2011vj}. More specifically, they provided a
compelling argument that in the flow from a UV CFT to an IR CFT the
inequality $a_{\text{UV}}>a_{\text{IR}}$ is satisfied, without however
providing a monotonically-decreasing $c$-function. Attempts to extend that
line of reasoning to the $d=6$ case have been unsuccessful, although in
explicit examples the validity of $a_{\text{UV}}>a_{\text{IR}}$ was
demonstrated~\cite{Elvang:2012st}. The method we use in this work is very
different from that used in~\cite{Komargodski:2011vj, Elvang:2012st}, and
allows us to obtain results local in the RG scale.

The organization of the paper is as follows. In the next section we
summarize the results of Osborn in 2d and 4d. We continue in Section
\ref{sec:Sixd} by describing the 6d case in detail, and we then illustrate
in Section~\ref{sec:Evend} the ingredients that allow us to genaralize our
analysis regarding the $a$-theorem to all even spacetime dimensions. In
Appendices~\ref{App:ConvDef} and~\ref{App:Terms} we present details
regarding our conventions as well as the terms that participate in the
consistency conditions in 6d. A Mathematica file that contains all the
consistency conditions in 6d is included with our submission.

\newsec{Summary of the 2d and 4d cases}
To begin, let us introduce the basic setting. More details can be found
in~\cite{Osborn:1991gm, Osborn:1991mk}. We are working in Euclidean space
and we define the generating functional $W$ of connected Green functions
via
\eqn{e^W=\int [d\phi]\,e^{-S},}
where $S$ is the Euclidean action with all required counterterms. $S$
contains a potential of the form $g^i\cO_i$, where $g^i$ are parameters
which can be taken to be dimensionless, and $\cO_i$ are
scaling-dimension-$d$ operators where $d$ is the spacetime
dimension.\foot{Non-marginal operators can also be included,
see~\cite{Osborn:1991gm}.} $W$ is a function of the renormalized couplings
$g^i$ and the metric $\gamma_{\mu\nu}$.

Consider now the RG flow as a flow in the space of theories as parametrized
by their couplings $g^i$.  The arbitrary RG parameter $\mu$ has to be
introduced, and the flow is then generated by
\eqn{\mathscr{D}=\mu\frac{\partial}{\partial \mu}
+\beta^i\frac{\partial}{\partial g^i},}
where $\beta^i=\mu\,dg^i/d\mu$ is the beta function. $W$ is a finite scalar
function, since it is derived from $S$ which includes all necessary
counterterms, and it is thus invariant under the RG flow:
\eqn{\mathscr{D}W=0.}[CSeq]
This is simply the Callan--Symanzik equation.

To define a local RG equation we let the parameters of $S$ as well as the
spacetime metric be arbitrary functions of spacetime. New counterterms
involving derivatives on the metric and the couplings are then necessary
for finiteness. With their inclusion in $S$ functional differentiations of
$W$ are guaranteed to produce finite operator-insertions in Green
functions.  Local rescalings of length are described by
\eqn{\gamma^{\mu\nu}(x)\to e^{2\sigma(x)}\gamma^{\mu\nu}(x)}
and they form the Weyl group.

We define the quantum stress-energy tensor and finite composite operators
using
\eqn{T_{\mu\nu}(x)=2\frac{\delta S}{\delta\gamma^{\mu\nu}(x)},\qquad
[\cO_i(x)]=\frac{\delta S}{\delta g^i(x)},}
where functional derivatives are defined in $d$ spacetime dimensions by
\eqn{\frac{\delta}{\delta\gamma^{\mu\nu}(x)}\gamma^{\kappa\lambda}(y)
=\tfrac12\delta_{\smash{(\mu}}^{\phantom{(\mu}\!\kappa}
\delta_{\smash{\nu)}}^{\phantom{\nu}\lambda}\,
\delta^{\hspace{0.5pt}d}(x,y),\qquad
\frac{\delta}{\delta g^i(x)}g^j(y)=\delta_i^{\phantom{i}\!j}\,
\delta^{\hspace{0.5pt}d}(x,y),}
with $X_{(i}Y_{j)}\equiv X_iY_j+X_jY_i$,
$\delta^{\hspace{0.5pt}d}(x,y)=\delta^{\hspace{0.5pt}d}(x-y)/\sqrt{\gamma(x)}$,
$\gamma$ is the determinant of the metric, and
$\delta^{\hspace{0.5pt}d}(x)$ the usual delta function in $d$ dimensions.
At the level of the generating functional we implement infinitesimal local
Weyl transformations with the generators
\eqn{\DeltaW_\sigma=2\int \dd x\sqrt{\gamma}\,\sigma\gamma^{\mu\nu}
\frac{\delta}{\delta\gamma^{\mu\nu}},\qquad \Delta_\sigma^\beta
=\int \dd x\sqrt{\gamma}\,\sigma\beta^i \frac{\delta}{\delta g^i}.}
With these definitions it is obvious that
\eqn{\DeltaW_\sigma W=-\int \dd x\sqrt{\gamma}\,\sigma\gamma^{\mu\nu}
\langle T_{\mu\nu}\rangle,\qquad
\Delta_\sigma^\beta W=-\int \dd x\sqrt{\gamma}\,\sigma\langle
\beta^i[\cO_i]\rangle.}
It is known that the Weyl variation of $W$, $\DeltaW_\sigma W$, is
anomalous in curved space~\cite{Capper:1974ic}, even when the flat-space
theory is a CFT.

In general, one can write
\eqn{\DeltaW_\sigma W=\Delta_\sigma^\beta W +\int
\dd x\sqrt{\gamma}\,(\text{terms with derivatives on
}\gamma_{\mu\nu},g^i,\sigma).}[VarW]
For a classically scale invariant theory we also have
\eqn{\left(\mu\frac{\partial}{\partial\mu}+ 2\int
\dd x\sqrt{\gamma}\,\gamma^{\mu\nu}\frac{\delta}{\delta\gamma^{\mu\nu}}
\right)W=0,}
and so if the integral in \VarW is neglected, then \VarW reduces to the
Callan--Symanzik equation \CSeq. Therefore, \VarW serves as a local version
of the Callan--Symanzik equation. It is straightforward to see that \VarW
is equivalent to
\eqn{\gamma^{\mu\nu}T_{\mu\nu}=\beta^i[\cO_i]+(\text{curvature,
}\partial_\mu g)\text{-terms}.}[TraceAnomaly]
This is the most general form of the trace anomaly.\foot{Actually there is
a more general form that includes renormalization effects of a specific
vector operator of classical scaling dimension $d-1$. In $d=4$ such
operators were considered by Osborn~\cite{Osborn:1991gm}, and were also
found in dimensional regularization in~\cite{Jack:1990eb}. For the
significance of such contributions the reader is referred
to~\cite{Fortin:2012hn}.} Consistency conditions follow from requiring that
$[\DeltaW_\sigma-\Delta_\sigma^\beta, \DeltaW_{\sigma^\prime}
-\Delta_{\sigma^\prime}^\beta]W=0$, as imposed by the fact that the Weyl
group is Abelian.

\subsec{The 2d case}
In two dimensions, the number of terms that are diffeomorphism and scale
invariant that can contribute to the trace anomaly is small, and the
elegance of the $c$-theorem is manifest. There is one curvature term and
two terms with derivatives on spacetime-dependent couplings one can write
down.  The trace anomaly is reproduced by
\eqn{\DeltaW_\sigma W=\Delta_\sigma^\beta W -\int \dtwo x\sqrt{\gamma}\,
\sigma(\tfrac12\beta^\Phi R-\tfrac12\chi_{ij}\partial_\mu
g^i\partial^\mu g^j) +\int \dtwo x\sqrt{\gamma}\,\partial_\mu\sigma\,
w_i\partial^\mu g^i,}[AnomTwod]
where $R$ is the Ricci scalar and $\beta^{\Phi}$, $\chi_{ij}$, and $w_i$
are functions of the couplings. From $[\DeltaW_\sigma-\Delta_\sigma^\beta,
\DeltaW_{\sigma^\prime}-\Delta_{\sigma^\prime}^\beta]W=0$ with \AnomTwod we
obtain a single consistency condition, namely
\eqn{\partial_\mu\beta^\Phi +
w_i\,\partial_\mu\beta^i +
\beta^i\partial_iw_j\,\partial_\mu g^j
-\chi_{ij}\beta^i\,\partial_\mu g^j=0,}
where $\partial_i\equiv\partial/\partial g^i$. Since this has to be true
for arbitrary $\partial_\mu g^i$, we conclude that
\eqn{\partial_i\tilde{\beta}^\Phi=\chi_{ij}\beta^j+\partial_{[i}w_{j]}
\beta^j,\qquad \tilde{\beta}^\Phi=\beta^\Phi+w_i\beta^i,}[cCCTwod]
where $X_{[i}Y_{j]}\equiv X_iY_j-X_jY_i$. By multiplying \cCCTwod by
$\beta^i$ we get,
\eqn{\frac{d\tilde{\beta}^\Phi}{dt}=-\chi_{ij}\beta^i\beta^j,}
which is equivalent to Zamolodchikov's $c$-theorem if $\chi_{ij}$ is
positive-definite.

One should note here that there is a degree of arbitrariness in the
definition of the various coefficients in \AnomTwod, corresponding to the
addition of allowed terms in the generating functional of the original 2d
theory. Indeed, if we shift $W\to W+\delta W$, where
\eqn{\delta W=\int \dtwo x\sqrt{\gamma}\,(\tfrac12 bR-\tfrac12
b_{ij}\partial_\mu g^i\partial^\mu g^j),}
with arbitrary functions $b$, $b_{ij}$, then
\eqna{\delta\beta^\Phi&=\beta^i\partial_ib=\mathscr{L}_\beta b,\qquad
\delta w_i=-\partial_ib+b_{ij}\beta^j\\
\delta\chi_{ij}&=\beta^k\partial_k b_{ij}
+\partial_i\beta^k\,b_{jk}+\partial_j\beta^k\,b_{ik}=\mathscr{L}_\beta
b_{ij},}
where $\mathscr{L}_\beta$ is the Lie derivative along the beta-function
vector.  Nevertheless, the consistency condition is invariant under this
arbitrariness.\foot{The arbitrariness we are discussing here is analogous
to the arbitrariness that affects the coefficient of $\square R$ in the 4d
trace anomaly at the fixed point.  In 2d we see that outside the fixed
point $\beta^\Phi$ has a degree of arbitrariness. Of course when
$\beta^i=0$ the well-defined $\beta^\Phi$ is the central charge of the
corresponding CFT (up to normalization).}  Osborn then establishes that
there is a choice of the arbitrariness so that the corresponding
$\chi_{ij}$ is positive-definite, essentially equal to Zamolodchikov's
metric $G_{ij}=(x^2)^2\left\langle[\cO_i(x)][\cO_j(0)] \right\rangle$. With
that choice $\tilde{\beta}^\phi$ becomes Zamolodchikov's $c$-function $C$.
As a final remark let us point out here that possible dimension-one vector
operators are neglected in the treatment of Osborn---such operators were
considered in~\cite{Friedan:2009ik}.

\subsec{The 4d case}
In four dimensions the elegance of the two-dimensional case is obfuscated
by the fact that there exist four curvature invariants (that conserve
parity) and quite a few terms that involve derivatives on the couplings.
The terms that account for the trace anomaly may be written as
\eqn{\DeltaW_\sigma W=\Delta_\sigma^\beta W+\int \dfour x
\sqrt{\gamma}\,\sigma\mathscr{T} +\int \dfour x\sqrt{\gamma}\,
\partial_\mu\sigma\,\mathscr{Z}^\mu,}[AnomFourd]
where
\eqna{\mathscr{T} &=\beta_aI+\beta_bE_4+\tfrac19\beta_cR^2
+\tfrac13\chi^e_i\partial_\mu g^i\partial^\mu R+\tfrac16
\chi^f_{ij}\partial_\mu g^i\partial^\mu g^j R +\tfrac12
\chi^g_{ij}\partial_\mu g^i\partial_\nu g^j
G^{\mu\nu}\\
&\quad +\tfrac12 \chi^a_{ij}\nabla^2 g^i\nabla^2
g^j+\tfrac12\chi^b_{ijk}\partial_\mu g^i\partial^\mu g^j\nabla^2 g^k
+\tfrac14 \chi^c_{ijkl}\partial_\mu g^i \partial^\mu
g^j \partial_\nu g^k\partial^\nu g^l,}[TFourd]
and
\eqna{\mathscr{Z}^\mu&=G^{\mu\nu}w_i\partial_\nu g^i +\tfrac13
\partial^\mu(b^\prime R)+\tfrac13 RY_i\partial^\mu g^i\\
&\quad +\partial^\mu(U_i\nabla^2 g^i+\tfrac12 V_{ij}\partial_\nu g^i
\partial^\nu g^j)+S_{ij}\partial^\mu g^i\nabla^2 g^j+\tfrac12
T_{ijk}\partial_\nu g^i\partial^\nu g^j\partial^\mu g^k,}[ZFourd]
up to terms with vanishing divergence. Definitions of the various
curvatures can be found in \eqref{BFour}. $G_{\mu\nu}$ is the Einstein
tensor and the various coefficients are functions of the couplings.

Here one finds six consistency conditions (which can be further
decomposed). Two of them are particularly interesting. First, there is a
consistency condition like \cCCTwod involving $\beta_b$:
\eqn{\partial_i\tilde{\beta}_b=\tfrac18\chi_{ij}^g\beta^j +\tfrac18
\partial_{[i}w_{j]}\beta^j,\qquad \tilde{\beta}_b=\beta_b +\tfrac18
w_i\beta^i.}[aCCFourd]
The consistency condition \aCCFourd can lead to an extension of
Zamolodchikov's result to 4d if the metric $\chi_{ij}^g$ can be shown to be
positive-definite. Of course, just like in 2d, there is an arbitrariness in
the definition of $\chi_{ij}^g$ as well as in other coefficients in \TFourd
and \ZFourd. To get an $a$-theorem it suffices to show that there is a
choice of the arbitrariness so that $\chi_{ij}^g$ is positive-definite.
This relies on the fact that \aCCFourd is invariant under the
arbitrariness.

The other consistency condition we would like to draw attention to is
\eqn{\beta_c=\tfrac14(\partial_ib^{\prime}-\chi_i^e)\beta^i.}
This shows that the coefficient of $R^2$ in the trace anomaly is generally
non-zero outside the fixed point. It also motivates the use of the term
``vanishing anomalies'' for contributions to the trace anomaly like $R^2$
in $d=4$: these are anomalies that are present along the flow but vanish at
the fixed point.

In our treatment so far we have neglected relevant operators with classical
scaling dimension three or two that may be present in a four-dimensional
theory. Osborn has considered such operators in~\cite{Osborn:1991gm}, and
has shown that the condition \aCCFourd is actually unaffected by their
presence, except for a shift of $\beta^i$ due to the presence of
dimension-three vector operators. This shift played an important role
in~\cite{Fortin:2012hn}, where it was calculated at three loops in the most
general renormalizable 4d QFT, and was used to show that at the
perturbative level scale implies conformal invariance in unitary
renormalizable 4d QFTs.

\newsec{The 6d case}[sec:Sixd]
As we saw in the previous section the elegance of the consistency
conditions rapidly disappears in the jump from 2d to 4d. Nevertheless, a
consistency condition similar to \cCCTwod remains, and it is interesting to
see if this is an accident or if such a consistency condition can be
obtained in higher (even) dimensions. This is the main motivation behind
this work, and the treatment of the highly nontrivial 6d case gives us
valuable intuition that actually applies to all even dimensions.  We
postpone the discussion of the general even-$d$ case until the next
section, and we turn now to the consistency conditions in $d=6$. Appendices
\ref{App:ConvDef} and \ref{App:Terms} contain information on conventions,
basis choices, as well as the terms that appear in the trace anomaly in 6d
away from the fixed point.

\subsec{Basis of curvature tensors}
It is clear from the complexity of the 4d case that the situation in 6d is
significantly more challenging. As a first step we have to classify the
curvature tensors that can be used in the anomaly terms. Of course terms
without curvatures also need to be considered.

To begin, note that for the various contributions to
$(\DeltaW_\sigma-\Delta_\sigma^\beta) W$ we are only constrained by
diffeomorphism invariance and power counting.  Let us look at a consequence
of this in 4d: in anomaly terms with one power of curvature one cannot
involve the Riemann tensor (without contracting its indices).  Indeed, the
Riemann tensor has four free indices, for which we would need four
derivatives on one or more couplings.  This would result in a term with
mass dimension six.  Therefore, in 4d, one can only include curvature
tensors with up to two free indices, and those are $R$, $\gamma_{\mu\nu}R$,
and $R_{\mu\nu}$.\foot{Incidentally, using the same argument one sees that
$\nabla_\lambda G_{\mu\nu}$, where $G_{\mu\nu}$ is the Einstein tensor, can
also not be included in the anomaly terms.} Since the variation of the 4d
Euler density in $d=4$ is
$\delta_\sigma(\sqrt{\gamma}E_4)=-8\sqrt{\gamma}\,G^{\mu\nu}
\nabla_\mu\partial_\nu \sigma$, it is preferable to include the Einstein
tensor instead of the Ricci tensor. This choice produces the consistency
conditions in a convenient form, but it is not essential.  Indeed, the
consistency conditions in a specific basis can be recast to the form
obtained in any other basis by a redefinition of the coefficients of the
various anomaly terms.

In 6d a similar choice is dictated by the fact that the Weyl variation of
the 6d Euler density is
\eqn{\delta_\sigma(\sqrt{\gamma}E_6)=12\sqrt{\gamma}\,(3E_4\gamma^{\mu\nu}
-2RR^{\mu\nu}+4R^{\mu}_{\hphantom{\mu}\!\kappa}R^{\kappa\nu}
+4R_{\kappa\lambda}R^{\kappa\mu\lambda\nu}
-2R_{\kappa\lambda\rho}^{\hphantom{\kappa\lambda\rho}\!\smash{\mu}}
R^{\kappa\lambda\rho\nu})\nabla_\mu\partial_\nu\sigma,}
where $E_4$ is given in even $d>2$ by $E_4=\tfrac{2}{(d-2)(d-3)}
(R^{\kappa\lambda\mu\nu}R_{\kappa\lambda\mu\nu}
-4R^{\kappa\lambda}R_{\kappa\lambda}+R^2)$. The tensors quadratic in
curvature that we have to consider can be found in \eqref{BFour}; the
tensor $H_{1}^{\mu\nu}$ is chosen so that in $d=6$ the variation of the
Euler density is $\delta_\sigma(\sqrt{\gamma}E_6)=
6\sqrt{\gamma}\,H_{1}^{\mu\nu} \nabla_\mu\partial_\nu\sigma$. As far as
terms quadratic in curvature are concerned, we also have to include the
terms \eqref{BFive}, which are basically derivatives of the terms in
\eqref{BFour}.  Terms linear in curvature include \eqref{BTwo} and
\eqref{BThree}. In writing down the various curvature tensors one has to
identify a complete but not over-complete basis, a problem complicated by
the symmetries of the Riemann tensor and the Bianchi identities.

As far as scalar terms cubic in curvature are
concerned~\cite{Bonora:1985cq, Bastianelli:2000rs}, the situation is
slightly more subtle. We have to include the terms in \eqref{BSix}, but
among them there are trivial anomalies, i.e.\ the terms $J_{1,\ldots,6}$
whose coefficient can be varied at will by a choice of local counterterms.
These are not genuine anomalies, but they nevertheless appear in the trace
anomaly, even at the fixed point.  The well-known example is the term
$\square R$ in 4d. In \eqref{BSix} there are also vanishing anomalies,
i.e.\ curvature terms that have to be included outside the fixed point, but
that do not satisfy the consistency conditions at the fixed point and thus
their coefficient has to be set to zero there. These are the terms
$L_{1,\ldots,7}$ in \eqref{BSix}. As we already mentioned there is only one
such term in 4d, namely $R^2$. Here, the form of $L_{1,\ldots,6}$ is chosen
based on the fact that these are the terms that shift the coefficients of
the trivial anomalies at the fixed point, i.e.\ $\delta_\sigma\int \dsix
x\sqrt{\gamma}\, L_{1,\ldots,6}=\int \dsix x\sqrt{\gamma}\,\sigma
J_{1,\ldots,6}$.

While $J_{1,\ldots,6}$ can be included in the basis of terms cubic in
curvature, there is a more convenient choice based on the fact that in
order to show that $\delta_\sigma\int \dsix x\sqrt{\gamma}\,
L_{1,\ldots,6}=\int \dsix x\sqrt{\gamma}\,\sigma J_{1,\ldots,6}$ one has to
integrate by parts.  But since total derivatives can be neglected in our
considerations (for $\sigma$ can be taken to have local support), this
implies that we don't have to include the trivial anomalies
$J_{1,\ldots,6}$ in $(\DeltaW_\sigma-\Delta_\sigma^\beta) W$, so long as we
include terms arising from $\delta_\sigma\int \dsix x\sqrt{\gamma}\,
z_{1,\ldots,6} L_{1,\ldots,6}$ before any integrations by parts. (Here
$z_{1,\ldots,6}$ are arbitrary functions of the couplings.) Consequently,
terms cubic in curvature that we need to consider are the three terms
$I_{1,2,3}$ that lead to Weyl-invariant densities, the 6d Euler term $E_6$,
and the seven vanishing anomalies $L_{1,\ldots,7}$. As we explained, this
relies on the ability to discard total derivatives.

\subsec{Contributions to the anomaly}
Now that we have a complete basis of curvature tensors we are ready to
write down the most general anomaly functional $(\DeltaW_\sigma
-\Delta_\sigma^\beta) W$. It takes the form
\eqn{\DeltaW_\sigma W = \Delta_\sigma^\beta W + \sum_{p=1}^{65}\int \dsix
x\sqrt{\gamma}\,\sigma\st_p+ \sum_{q=1}^{30}\int \dsix
x\sqrt{\gamma}\,\partial_\mu\sigma\,\sz^\mu_q,}
where the $\st_p$ and $\sz_q^\mu$ are dimension-six and dimension-five
terms respectively, that can involve curvatures as well as derivatives on
the couplings $g^i$ (see Appendix \ref{App:Terms}).

Much like Osborn did in $d=2,4$ we split the anomaly contributions into
terms with $\sigma$ and $\partial_\mu\sigma$. This splitting may seem
mysterious---as we could also introduce terms of the form $\int \dsix
x\sqrt{\gamma}\, \square\sigma\mathscr{V}$, for example---but it is used
here in order to get the consistency conditions in a most convenient form.
We can obtain the desired form of the consistency conditions even without
the splitting, if we carefully choose the coefficients of the various terms
in the anomaly.  This can be seen by integrating by parts to rewrite the
$\sz^\mu_q$ terms in the form of the $\st_p$ terms, which would lead to
some new $\st_p$ terms but also to shifts of coefficients of existing
$\st_p$ terms.

Let us illustrate this point more clearly in the 2d case.  Suppose that
instead of \AnomTwod we started with the equivalent
\eqn{\DeltaW_\sigma W=\Delta_\sigma^\beta W -\int \dtwo x\sqrt{\gamma}\,
\sigma(\tfrac12\beta^\Phi R-\tfrac12\chi^\prime_{ij}\partial_\mu
g^i\partial^\mu g^j + w_i\square g^i).}[OtherAnomTwod]
After an integration by parts of the $\square g^i$ term this amounts simply
to the definition $\chi_{ij}=\chi_{ij}^\prime+2\partial_{(i}w_{j)}$ in
\AnomTwod. This can also be seen by computing the Weyl consistency
condition from \OtherAnomTwod directly. We get
\eqn{\partial_i\tilde{\beta}^\Phi=(\chi_{ij}^\prime+2\partial_iw_j)\beta^j=
(\chi_{ij}^\prime+\partial_{(i}w_{j)})\beta^j+\partial_{[i}w_{j]}\beta^j.
}[OthercCCTwod]
Clearly, \OthercCCTwod is equivalent to \cCCTwod with the proper definition
of $\chi_{ij}$.

\subsec{Some consistency conditions}
Here we include some consistency conditions and we comment on the most
interesting ones. A Mathematica file with all the consistency conditions is
included with our submission.

Just like in 2d and 4d we obtain consistency conditions simply by the
requirement $[\DeltaW_\sigma-\Delta_\sigma^\beta,
\DeltaW_{\sigma^\prime}-\Delta_{\sigma^\prime}^\beta]W=0$. In our case we
find a total of forty one consistency conditions.\foot{Some of these
consistency conditions can be further decomposed as a result of the variety
of ways with which spacetime derivatives can act on couplings.} For
example, consistency requires that terms proportional to
$\partial_\mu\sigma\,\square^2\sigma^\prime-\partial_\mu\,\sigma^\prime
\square^2\sigma$ add up to zero, which leads to
\eqn{\partial_\mu(4b_{11}-3\mathcal{A}_i\beta^i)
+(4\mathcal{A}_i+2\mathcal{G}_i^4+5\cH_i^5
+2\cH_i^6-2\mathcal{I}^4_i)\,\partial_\mu
g^i+6\mathcal{A}_i\,\partial_\mu\beta^i
-\mathcal{A}_{ij}^\prime\beta^j\,\partial_\mu g^i=0,}
which implies that
\eqn{\partial_i(4b_{11}-3\mathcal{A}_j\beta^j)
+4\mathcal{A}_i+2\mathcal{G}_i^4+5\cH_i^5+2\cH_i^6
-2\mathcal{I}^4_i+6\mathcal{A}_j\,\partial_i\beta^j=
\mathcal{A}_{ij}^\prime\beta^j.}

Among the forty one consistency conditions in 6d the most interesting is
the one similar to \cCCTwod, obtained from terms proportional to
$(\sigma\partial_\mu\sigma^\prime-\sigma^\prime\partial_\mu\sigma)
H_1^{\mu\nu}$. It reads
\eqn{\partial_\mu(6a+b_1-\tfrac{1}{15}b_3) +
\cH^1_i\,\partial_\mu\beta^i +
\beta^i\partial_i\cH^1_j\,\partial_\mu g^j
- \cH_{ij}^1\beta^i\,\partial_\mu g^j=0,}
which can be brought to the form
\eqn{\partial_i\tilde{a}=\tfrac16\cH_{ij}^1\beta^j
+\tfrac16\partial_{[i}\cH^1_{j]}\beta^j, \qquad
\tilde{a}=a+\tfrac16 b_1-\tfrac{1}{90}b_3+\tfrac16\cH^1_i\beta^i.
}[aCCSixd]

The consistency condition \aCCSixd has a new feature compared to the 2d and
4d cases, i.e.\ that the function $\tilde{a}$ contains the coefficients
$b_1$ and $b_3$ of the vanishing anomalies $L_1$ and $L_3$ respectively.
This is of no consequence as far as the value of $\tilde{a}$ at the fixed
point is concerned:  there $\tilde{a} = a$, for $b_1=b_3=0$ at the fixed
point.  This fact is actually made explicit by three consistency
conditions.  More specifically, from terms proportional to
$(\partial_\kappa\sigma\,\nabla_\lambda\partial_\mu\sigma^\prime -
\partial_\kappa\sigma^\prime\,\nabla_\lambda\partial_\mu\sigma)
\nabla^\kappa G^{\lambda\mu}$, $(\sigma\partial_\mu \sigma^\prime-
\sigma^\prime\partial_\mu \sigma)\nabla_\nu H_4^{\mu\nu}$, and
$(\sigma\partial_\mu \sigma^\prime- \sigma^\prime\partial_\mu
\sigma)\nabla_\nu H_3^{\mu\nu}$ we find
\threeseqn{b_7&=\tfrac18\mathcal{F}_i\beta^i,}[DEinstein]{3b_1-8b_7&=
-\tfrac14(\partial_ib_{14}+\mathcal{I}_i^7)\beta^i,}[DHFour]{12b_1-b_3
-16b_7&=-(\partial_ib_{13}+\mathcal{I}_i^6)\beta^i,}[DHThree][DD]
respectively. From similar consistency conditions we can verify that $b_2,
b_4, b_5$ and $b_6$ are also zero at the fixed point, as expected since
they are coefficients of vanishing anomalies.

\subsec{Possibility for an \texorpdfstring{$a$}{a}-theorem in 6d}
The consistency condition \aCCSixd has the potential to lead to a result
similar to that of Zamolodchikov in 2d. Indeed, contracting with the beta
function it follows that \aCCSixd implies
\eqn{\frac{d\tilde{a}}{dt}=-\tfrac16\cH^1_{ij}\beta^i\beta^j.}[aThmSixd]
Note that here the conditions \DD allow us to absorb the $b_1$ and $b_3$
contributions in $\tilde{a}$ to a shift of $\cH^1_i$. Of course
what is missing is a proof of the positive-definiteness of
$\cH^1_{ij}$.

It is important to point out that the consistency condition \aCCSixd is
actually stronger than \aThmSixd. Indeed, \aCCSixd also contains
information about the possibility of a gradient flow interpretation of the
RG flow. For that, it has to be that $\partial_{[i}\cH^1_{j]}=0$,
in which case $\tilde{a}$ is the ``potential'' whose gradient produces the
RG flow.

Let us now concentrate on a technical but important point. It turns out
that the tensor $H_{1}^{\mu\nu}$, which appears in
$\delta_\sigma(\sqrt{\gamma}E_6)= 6\sqrt{\gamma}\,
H_{1}^{\mu\nu}\nabla_\mu\partial_\nu\sigma$, is divergenceless. A similar
statements holds in two,
$\delta_\sigma(\sqrt{\gamma}R)=2\sqrt{\gamma}\,\gamma^{\mu\nu}
\nabla_\mu\partial_\nu \sigma$, and four dimensions,
$\delta_\sigma(\sqrt{\gamma}E_4)=-8\sqrt{\gamma}\,G^{\mu\nu}
\nabla_\mu\partial_\nu\sigma$. This is actually crucial for the coefficient
of the Euler term to be involved in a consistency condition like \aCCSixd,
which has the chance to lead to an $a$-theorem. This is not so easy to see
in 2d and 4d, but it is clear in 6d.

Indeed, consider, for example, the consistency condition arising from
terms proportional to $(\sigma\partial_\mu\sigma^\prime-
\sigma^\prime\partial_\mu\sigma) H_4^{\mu\nu}$. It reads
\eqn{\partial_i \tilde{b}_1=\tfrac{1}{12}(\cH_{ij}^4
+\tfrac12\mathcal{F}_{ij})\beta^j +\tfrac{1}{12}\partial_{[i}
\cH_{j]}^4\beta^j +\tfrac16\mathcal{I}^7_i,\qquad
\tilde{b}_1=-b_1+\tfrac23b_7+\tfrac{1}{12}\cH^4_i\beta^i.
}[boneCCSixd]
The contribution $\tfrac16\mathcal{I}^7_i$ does not allow \boneCCSixd as a
candidate for the generalization of Zamolodchikov's result.\foot{Of course
this can also be seen from the fact that $\tilde{b}_1$ becomes zero at
fixed points, and so it cannot possibly be monotonically-decreasing along
an RG flow.} This contribution in fact arises from the term
$\st_{18}=\mathcal{I}^7_i\,\partial_\mu g^i\,\nabla_\nu H_4^{\mu\nu}$. Were
$\nabla_\nu H_1^{\mu\nu}$ non-vanishing, we would not be able to find a
consistency condition like \aCCSixd. It can be verified by explicit
computations that $H_1^{\mu\nu}$ is the only divergenceless symmetric
two-index tensor quadratic in curvature. It is thus a generalization of the
Einstein tensor.  As we will see in the next section Lovelock has
constructed all such generalizations a long time
ago~\cite{Lovelock:1971yv}, something that will allow us to argue for a
consistency condition similar to \aCCSixd in all even $d$.

\subsec{Arbitrariness}
Just like the coefficient of $\square R$ in the four-dimensional trace
anomaly, the various coefficients in $\st_p$ and $\sz_q^\mu$ are affected
by the choice of additive, quantum-field-independent counterterms. Indeed,
calculations in curved space and with $x$-dependent couplings will result
in infinities that will need to be renormalized via counterterms whose
finite part is arbitrary.  Therefore, different subtraction schemes will
result in different coefficients for $\st_p$ and $\sz_q^\mu$.

The most general addition to the generating functional of our theory is
\eqn{\delta W=-\sum_{p=1}^{65}\int \dsix x\sqrt{\gamma}\,\mathscr{X}_p,}
where the $\mathscr{X}_p$ terms have the same form as the $\st_p$ terms but
with arbitrary coefficients. There is no arbitrariness introduced by terms
$\mathscr{X}_q^\mu$ similar in form to the $\sz_q^\mu$ terms, for those are
total derivatives. Now, the consistency conditions are invariant under the
shift $W\to
W+\delta W$, although the coefficients in the consistency
conditions will shift. Let us see how this works for \aCCSixd.

The relevant terms are
\eqn{\int \dsix x\sqrt{\gamma}\,(z_aE_6+z_{b_1}L_1+z_{b_3}L_3
-\tfrac12 z_{ij}^{\cH^1}\partial_\mu g^i\partial_\nu g^j
H_1^{\mu\nu})\subset\delta W}[ArbSixd]
and one can verify that their inclusion leads to shifts
\eqna{\delta a&=\mathscr{L}_\beta z_a,\qquad
\delta b_1=\mathscr{L}_\beta z_{b_1},\qquad
\delta b_3=\mathscr{L}_\beta z_{b_3},\\
\delta \cH^1_i&=-6\partial_i(z_a+\tfrac16
z_{b_1}-\tfrac{1}{90}z_{b_3})+z_{ij}^{\cH^1}\beta^j,\qquad
\delta\cH^1_{ij}=\mathscr{L}_\beta
z_{ij}^{\cH^1},}[ArbShiftsSixd]
under which \aCCSixd is invariant. Note that $a$, which is of course
well-defined at the fixed point, is arbitrary along the flow, while
$\cH_i^1$ and $\cH_{ij}^1$ have a degree of arbitrariness
even at the fixed point. Also note that the shifts \ArbShiftsSixd cannot be
used to set the corresponding coefficients to zero, except for
$\cH_i^1$ if $\cH_{i}^1=\partial_i X$ for some $X$.

This observation leads to an important point, which we have already
emphasized: regarding the $a$-theorem, one should not be able to prove that
the metric $\cH_{ij}^1$ is positive-definite in all generality.
Instead, one ought to be able to show that there is a choice for the
arbitrariness \ArbSixd such that $\cH_{ij}^1$ is positive-definite.
That specific choice then gives us the quantity $\tilde{a}$ whose flow is
monotonic, through the dependence of $\delta\cH_i^1$ on
$z_{ij}^{\cH^1}$. Recall that in 2d arbitrariness similar to the
one described here was used by Osborn to rederive Zamolodchikov's
$c$-theorem (see~\cite{Osborn:1991gm} for details).

\newsec{Consistency conditions in even spacetime dimensions}[sec:Evend]
In this section we identify the ingredients that allow us to conclude that
a consistency condition like \aCCSixd appears in all even spacetime
dimensions. Of course non-trivial CFTs in $d>6$ are not known, but it is
still interesting to consider the generalization of our results.

According to the classification of~\cite{Deser1993279}, for a CFT in any
even spacetime dimension lifted to curved space the conformal anomaly
consists of a unique Euler term (type-A anomaly), a number of terms that
lead to locally Weyl invariant densities (type-B anomalies), as well as a
number of trivial anomalies. Outside the fixed point we also have a number
of vanishing anomalies. As for the trivial anomalies, these can always be
accounted for by terms with $d/2-1$ powers of curvature.

Now, in any even spacetime dimension, $d=2n$, it is easy to see that the
Weyl variation of the Euler density $\sqrt{\gamma}E_{2n}$, where
\eqn{E_{2n}=\frac{1}{2^n}R_{i_1j_1k_1l_1}\cdots R_{i_nj_nk_nl_n}
\epsilon^{i_1j_1\ldots i_nj_n}\epsilon^{k_1l_1\ldots k_nl_n},}
gives
\eqn{\delta_\sigma(\sqrt{\gamma}E_{2n})=
\sqrt{\gamma}\,H^{\mu\nu}\nabla_\mu\partial_\nu\sigma,}
for some symmetric tensor $H^{\mu\nu}$ with $n-1$ powers of the curvature.
As Lovelock showed~\cite{Lovelock:1971yv}, this tensor $H^{\mu\nu}$ is the
unique tensor with the properties of the Einstein tensor---in particular,
it is the only two-index symmetric tensor with $n-1$ powers of the
curvature that is divergenceless:
\eqn{\nabla_\nu H^{\mu\nu}=0.}
Regarding the consistency condition similar to \aCCSixd, this observation
allows us to conclude that the only relevant terms among the various
contributions to the anomaly $(\DeltaW_\sigma-\Delta_\sigma^\beta)W$ are
\eqn{\int d^{\hspace{0.5pt}2n}\hspace{-0.5pt}x\sqrt{\gamma}\,
\sigma\left[(-1)^naE_{2n}+\sum_p b_pL_p
+\tfrac12\cH_{ij}\,\partial_\mu g^i\partial_\nu g^j\,
H^{\mu\nu}\right]+\int d^{\hspace{0.5pt}2n}\hspace{-0.5pt}x\sqrt{\gamma}\,
\partial_\mu\sigma\,\cH_i\,\partial_\nu g^i\,H^{\mu\nu},}
where $L_p$ are some vanishing anomalies. A consistency condition similar
to \aCCSixd is thus easily found, and is of course invariant under
arbitrariness generated by contributions similar to \ArbSixd.

A relation of the metric $\cH_{ij}$ to a positive-definite metric
is currently only known in 2d~\cite{Osborn:1991gm}. A similar relation in
higher even $d$ is lacking, but its possible existence would immediately
imply the generalization of Zamolodchikov's result. To summarize, in any
even spacetime dimension one can find a scalar quantity $\tilde{a}$ such
that
\eqn{\partial_i\tilde{a}=\cH_{ij}\beta^j
+\partial_{[i}\cH_{j]}\beta^j.}[aCC]
The quantity $\tilde{a}$ becomes the coefficient of the Euler term in the
trace anomaly at the fixed point, but more generally it includes a linear
combination of the $b_p$s and a term $\cH_i\beta^i$. The relation \aCC
immediately implies that
\eqn{\frac{d\tilde{a}}{dt}=-\cH_{ij}\beta^i\beta^j,}
which, if $\cH_{ij}$ can be related to a positive-definite metric
via the arbitrariness $\delta\cH_{ij}=\mathscr{L}_\beta
z_{ij}^{\cH}$ with $z_{ij}^\cH$ an arbitrary symmetric tensor, is the
generalization of the 2d $c$-theorem.

\ack{We have relied heavily on Mathematica and the package
\href{http://www.xact.es/}{\texttt{xAct}}. We would like to thank Aneesh
Manohar, John McGreevy, and especially Ken Intriligator for helpful
discussions.  This work was supported in part by the US Department of
Energy under contract DE-SC0009919.}

\appendix{}

\newsec{Conventions and definitions}[App:ConvDef]
Throughout this paper we follow the conventions of Misner, Thorne and
Wheeler~\cite{Misner:1974qy} for the Riemann tensor. For the Weyl variation
of the metric we choose
\eqn{\gamma_{\mu\nu}\to e^{-2\sigma}\gamma_{\mu\nu}.}
Infinitesimally, then,
$\delta_\sigma\gamma_{\mu\nu}=-2\sigma\gamma_{\mu\nu}$ and so
$\delta_\sigma\gamma^{\mu\nu}=2\sigma\gamma^{\mu\nu}$ (we do not use
$\delta\sigma$ for an infinitesimal $\sigma$, since no confusion can
arise).

It is important to classify the curvature terms of various mass dimensions.
These will be used subsequently to construct all possible terms that can
appear in $(\DeltaW_\sigma-\Delta_\sigma^\beta)W$. In two and four
spacetime dimensions this is very easy, but in six it becomes a rather
cumbersome problem, plagued by complications due to the large number of
monomials and the identities of the Riemann tensor.

A complete basis $\mathfrak{B}_2$ of dimension-two curvature terms that can
be used in $\DeltaW_\sigma W$ is given by the Ricci scalar, the Einstein
tensor, and the Riemann tensor,
\eqn{\tfrac{1}{d-1}R,\qquad  G_{\mu\nu},\qquad
R_{\kappa\lambda\mu\nu},}[BTwo]
where we define the Einstein tensor as
\eqn{G_{\mu\nu}=\tfrac{2}{d-2}(R_{\mu\nu}-\tfrac12 \gamma_{\mu\nu}R)\qquad
(d\ge3),}
where $R_{\mu\nu}$ is the Ricci tensor.  Taking a derivative leads to three
dimension-three terms, but, by diffeomorphism invariance and simple power
counting, only two can be used in $\DeltaW_\sigma W$, namely
\eqn{\tfrac{1}{d-1}\partial_\mu R\qquad \text{and}\qquad \nabla_\kappa
G_{\mu\nu}.}[BThree]
These form the basis $\mathfrak{B}_3$.

At the level of dimension-four curvature terms only terms with up to two
free indices are allowed in $\DeltaW_\sigma W$. We consider the basis
$\mathfrak{B}_4$ with elements
\eqn{\begin{gathered}
E_4=\tfrac{2}{(d-2)(d-3)}(R^{\kappa\lambda\mu\nu}R_{\kappa\lambda\mu\nu}
  -4R^{\kappa\lambda}R_{\kappa\lambda}+R^2),\\
I=R^{\kappa\lambda\mu\nu}R_{\kappa\lambda\mu\nu}
  -\tfrac{4}{d-2}R^{\kappa\lambda}R_{\kappa\lambda}
  +\tfrac{2}{(d-1)(d-2)} R^2,\qquad
\tfrac{1}{(d-1)^2}R^2,\qquad
\tfrac{1}{d-1}\square R,\\
H_{1\mu\nu}=\tfrac{(d-2)(d-3)}{2}E_4\gamma_{\mu\nu}-4(d-1)H_{2\mu\nu}
  +8H_{3\mu\nu}+8H_{4\mu\nu}
  -4R^{\kappa\lambda\rho}_{\hphantom{\kappa\lambda\rho}\!\mu}
  R_{\kappa\lambda\rho\nu},\\
H_{2\mu\nu}=\tfrac{1}{d-1}RR_{\mu\nu},\qquad
H_{3\mu\nu}=R_{\mu}^{\hphantom{\mu}\!\kappa}R_{\kappa\nu},\qquad
H_{4\mu\nu}=R^{\kappa\lambda}R_{\kappa\mu\lambda\nu},\\
H_{5\mu\nu}=\square R_{\mu\nu},\qquad
H_{6\mu\nu}=\tfrac{1}{d-1}\nabla_\mu\partial_\nu R.
\end{gathered}}[BFour]
All $H_{1,\ldots,6}$ are symmetric. $I$ is the Weyl tensor squared,
$I=W^{\kappa\lambda\mu\nu}W_{\kappa\lambda\mu\nu}$, and $\sqrt{\gamma}E_4$
is the four-dimensional Euler density. In our conventions the Weyl tensor
is given by
\eqn{W_{\kappa\lambda\mu\nu}=R_{\kappa\lambda\mu\nu}+
\tfrac{2}{d-2}(\gamma_{\kappa[\nu}R_{\mu]\lambda}+
\gamma_{\lambda[\mu}R_{\nu]\kappa})+
\tfrac{2}{(d-1)(d-2)}\gamma_{\kappa[\mu}\gamma_{\nu]\lambda}R \qquad
(d\ge3).}

The dimension-five curvature terms we need to consider are given by
\eqn{\begin{gathered}
\partial_\mu E_4,\qquad
\partial_\mu I,\qquad
\tfrac{1}{(d-1)^2}R\,\partial_\mu R,\qquad
\tfrac{1}{d-1}\partial_\mu\square R,\qquad
\nabla^\nu H_{(2,3,4)\mu\nu},
\end{gathered}}[BFive]
and they form the basis $\mathfrak{B}_5$. Note that we do not need
$\nabla^\nu H_{1\mu\nu}$, for $\nabla^\nu H_{1\mu\nu}=0$.\foot{In any even
dimension $d$, the Weyl variation of $\sqrt{\gamma}E_d$ is
$\sqrt{\gamma}\,H^{\mu\nu}\nabla_\mu\partial_\nu\sigma$ for some symmetric
tensor $H^{\mu\nu}$. Since $\delta_\sigma\int \dd  x\sqrt{\gamma}E_d=0$, it
follows that $\nabla_\mu\nabla_\nu H^{\mu\nu}=0$. In \cite{Lovelock:1971yv}
it was shown, however, that $H^{\mu\nu}$ is actually divergenceless, i.e.\
$\nabla_\nu H^{\mu\nu}=0$.}  Similarly, $\nabla^\nu H_{(5,6)\mu\nu}$ are
not necessary, for
\eqn{\nabla^\nu H_{5\mu\nu}=\nabla^\nu\{(d-1)H_{2\mu\nu}-
2H_{4\mu\nu} -\tfrac12\gamma_{\mu\nu}[\tfrac18(d-2)^2E_4
-\tfrac{d-2}{4(d-3)} I +\tfrac{d-2}{4(d-1)}R^2 -
\square R]\}}
and
\eqn{\nabla^\nu H_{6\mu\nu}=\nabla^\nu[H_{2\mu\nu}
-\tfrac{1}{d-1}\gamma_{\mu\nu}(\tfrac14 R^2 -\square R)].}
The corresponding matrix of coefficients of the remaining terms in
$\nabla^\mu\mathfrak{B}_4$ has full rank, which shows that $\mathfrak{B}_5$
is a good basis.

Finally, a complete basis of scalar dimension-six curvature terms was
constructed in \cite{Bonora:1985cq}. Its building blocks are
$K_{1,\ldots,17}$ given by
\eqn{\begin{gathered}
K_1=R^3,\qquad
K_2=RR^{\kappa\lambda}R_{\kappa\lambda},\qquad
K_3=RR^{\kappa\lambda\mu\nu}R_{\kappa\lambda\mu\nu},\qquad
K_4=R^{\kappa\lambda}R_{\lambda\mu}R^\mu_{\hphantom{\mu}\!\kappa},\\
K_5=R^{\kappa\lambda}R_{\kappa\mu\nu\lambda}R^{\mu\nu},\qquad
K_6=R^{\kappa\lambda}R_{\kappa\mu\nu\rho}
  R_{\lambda}^{\smash{\hphantom{\lambda}\mu\nu\rho}},\qquad
K_7=R^{\smash{\kappa\lambda\mu\nu}}R_{\mu\nu\rho\sigma}
  R^{\rho\sigma}_{\smash{\hphantom{\rho\sigma}\!\kappa\lambda}},\\
K_8=R^{\kappa\lambda\mu\nu}R_{\rho\lambda\mu\sigma}
  R_{\smash{\kappa\hphantom{\rho\sigma}\!\nu}}
  ^{\smash{\hphantom{\kappa}\rho\sigma}},\qquad
K_9=R\,\square R,\qquad
K_{10}=R^{\kappa\lambda}\,\square R_{\kappa\lambda},\qquad
K_{11}=R^{\kappa\lambda\mu\nu}\,\square R_{\kappa\lambda\mu\nu},\\
K_{12}=R^{\kappa\lambda}\nabla_\kappa\partial_\lambda R,\qquad
K_{13}=\nabla^\kappa R^{\lambda\mu}\nabla_{\kappa}R_{\lambda\mu},\qquad
K_{14}=\nabla^\kappa R^{\lambda\mu}\nabla_{\lambda}R_{\kappa\mu},\\
K_{15}=\nabla^\kappa R^{\lambda\mu\nu\rho} \nabla_\kappa
R_{\lambda\mu\nu\rho},
  \qquad
K_{16}=\square R^2,\qquad
K_{17}=\square^2 R.
\end{gathered}}
At the fixed point we can express the trace anomaly in the basis $K$'s, and
the consistency condition implies that there are seven combinations of
$K$'s whose coefficient has to be set to zero~\cite{Bonora:1985cq,
Bastianelli:2000rs}.  Thus, we can arrange the $K$'s in the basis
of~\cite{Bastianelli:2000rs},
\eqna{I_1&=\tfrac{19}{800}K_1 - \tfrac{57}{160}K_2 + \tfrac{3}{40}K_3 +
\tfrac{7}{16}K_4 - \tfrac{9}{8}K_5 - \tfrac{3}{4}K_6 +
K_8,\displaybreak[0]\\
I_2&=\tfrac{9}{200}K_1 - \tfrac{27}{40}K_2 + \tfrac{3}{10}K_3 +
\tfrac{5}{4}K_4 - \tfrac{3}{2}K_5 - 3K_6 + K_7,\displaybreak[0]\\
I_3&=-\tfrac{11}{50}K_1 + \tfrac{27}{10}K_2 - \tfrac{6}{5}K_3 - K_4 + 6K_5+
2K_7 - 8K_8\\&\hspace{3cm} + \tfrac{3}{5}K_9 - 6K_{10} + 6K_{11} + 3K_{13}
- 6K_{14} + 3K_{15},\displaybreak[0]\\
E_6&=K_1 - 12K_2 + 3K_3 + 16K_4 - 24K_5 - 24K_6 + 4K_7 +
8K_8,\displaybreak[0]\\
J_1&=6K_6 - 3K_7 + 12K_8 + K_{10} - 7K_{11} - 11K_{13} + 12K_{14} -
  4K_{15},\displaybreak[0]\\
J_2&=-\tfrac15 K_9 + K_{10} + \tfrac25 K_{12} + K_{13},\displaybreak[0]\\
J_3&=K_4 + K_5 - \tfrac{3}{20}K_9 + \tfrac45 K_{12}
  + K_{14},\displaybreak[0]\\
J_4&=-\tfrac15 K_9 + K_{11} + \tfrac25 K_{12} + K_{15},\displaybreak[0]\\
J_5&=K_{16},\displaybreak[0]\\
J_6&=K_{17},}
which makes manifest the splitting of anomalies at the fixed point into
type A ($E_6$) and B ($I_{1,2,3}$) according to the classification
of~\cite{Deser1993279}, and also trivial ($J_{1,\ldots,6}$). To be more
specific, $I_{1,2,3}$, which can be expressed as
\eqna{I_1&=W^{\kappa\lambda\mu\nu}W_{\rho\lambda\mu\sigma}
  W_{\smash{\kappa\hphantom{\rho\sigma}\!\nu}}
  ^{\smash{\hphantom{\kappa}\!\rho\sigma}},\\
I_2&=W^{\smash{\kappa\lambda\mu\nu}}W_{\mu\nu\rho\sigma}
  W^{\rho\sigma}_{\smash{\hphantom{\rho\sigma}\!\kappa\lambda},}\\
I_3&=W^{\kappa\lambda\mu\nu}(\delta_{\kappa}
  ^{\hphantom{\kappa}\!\rho}\,\square + 4R_{\kappa}
  ^{\hphantom{\kappa}\!\rho} -\tfrac65\delta_{\kappa}
  ^{\hphantom{\kappa}\!\rho}R)
  W_{\rho\lambda\mu\nu} - \tfrac23 J_1 -\tfrac{13}{3}J_2 + 2J_3
  + \tfrac13 J_4,}
lead to locally Weyl-invariant densities, while $J_{1,\ldots,6}$ can be set
to zero in the trace anomaly by a choice of local counterterm, just like
$\square R$ in four dimensions. The piece $-\tfrac23 J_1 -\tfrac{13}{3}J_2
+ 2J_3 + \tfrac13 J_4$ in $I_3$ is necessary for $\sqrt{\gamma}I_3$ to be
locally Weyl-invariant~\cite{Bastianelli:2000rs}.

For our purposes all $K$'s are needed, since we are interested in
consistency conditions valid along the RG flow. It is convenient to work in
the basis $\mathfrak{B}_6$ given by
\eqn{\begin{gathered}
I_1,\qquad
I_2,\qquad
I_3,\qquad
E_6,\\
L_{1}=-\tfrac{1}{30}K_1+\tfrac14 K_2-K_6,\qquad
L_{2}=-\tfrac{1}{100}K_1+\tfrac{1}{20}K_2,\\
L_{3}=-\tfrac{37}{6000}K_1+\tfrac{7}{150}K_2 -\tfrac{1}{75}K_3
+\tfrac{1}{10}K_5+\tfrac{1}{15}K_6,\qquad
L_{4}=-\tfrac{1}{150}K_1+\tfrac{1}{20}K_3,\\
L_{5}=\tfrac{1}{30}K_1,\qquad
L_{6}=-\tfrac{1}{300}K_1+\tfrac{1}{20}K_9,\qquad
L_{7}=K_{15},\\
J_1,\qquad
J_2,\qquad
J_3,\qquad
J_4,\qquad
\tfrac{1}{25}J_5,\qquad
\tfrac{1}{25}J_{6}.
\end{gathered}}[BSix]
The form of $L_{1,\ldots,6}$ is chosen based on the fact that these are the
terms that shift the coefficients of the trivial anomalies at the fixed
point, i.e.\ $\delta_\sigma\int \dsix x\sqrt{\gamma}\,L_{1,\ldots,6}=\int
\dsix x\sqrt{\gamma}\,\sigma J_{1,\ldots,6}$.

The choice of bases is arbitrary, and the form of the consistency
conditions depends on the choice. Nevertheless, the essential conclusions
derived from the consistency conditions are basis-independent.

\newsec{Terms in the anomaly}[App:Terms]
In six spacetime dimensions there are ninety five independent terms that
can contribute to $(\DeltaW_\sigma- \Delta_\sigma^\beta)W$.  We include
them in this appendix for easy reference.

In general, we can write
\eqn{\DeltaW_\sigma W = \Delta_\sigma^\beta W + \sum_{p=1}^{65}\int \dsix
x
\sqrt{\gamma}\,\sigma\st_p
+ \sum_{q=1}^{30}\int \dsix
x\sqrt{\gamma}\,\partial_\mu\sigma\,\sz^\mu_q.}
Clearly, the $\st_p$ and $\sz_q^\mu$ are dimension-six and dimension-five
terms respectively, that can involve curvatures as well as derivatives on
the couplings $g^i$. In writing down the various terms below, we neglect
total derivatives, and we keep in mind the convenient form in which we want
to obtain the consistency conditions.

If only curvatures are included, then we have the terms
\eqn{\begin{gathered}
\st_1=-c_1I_1,\qquad
\st_2=-c_2I_2,\qquad
\st_3=-c_3I_3,\qquad
\st_4=-aE_6,\qquad
\st_{5,\ldots,11}=-b_{1,\ldots,7}L_{1,\ldots,7}.
\end{gathered}}
We call these the $(0,6)$ terms, for only curvatures and derivatives on
curvatures contribute to the power counting. We also have $(0,5)^\mu$
terms, given by
\eqn{\begin{gathered}
\sz_1^\mu=-b_8\,\partial^\mu E_4,\qquad
\sz_2^\mu=-b_9\,\partial^\mu I,\qquad
\sz_3^\mu=-\tfrac{1}{25}b_{10}\,R\,\partial^\mu R,\\
\sz_4^\mu=-\tfrac15 b_{11}\,\partial^\mu\square R,\qquad
\sz_{5,6,7}^\mu=-b_{12,13,14}\,\nabla_\nu H_{2,3,4}^{\mu\nu}.
\end{gathered}}
Next, we can allow one power of $\partial_\mu g^i$ to get
\eqn{\begin{gathered}
\st_{12}=\cI_i^1\,\partial_\mu g^i\,\partial^\mu E_4,\qquad
\st_{13}=\cI_i^2\,\partial_\mu g^i\,\partial^\mu I,\qquad
\st_{14}=\tfrac{1}{25}\cI_i^3\,\partial_\mu g^i\,R\,\partial^\mu R,\\
\st_{15}=\tfrac{1}{5}\cI_i^4\,\partial_\mu g^i\,\partial^\mu\square R\qquad
\st_{16,17,18}=\cI_i^{5,6,7}\,\partial_\mu g^i\,\nabla_\nu
  H_{2,3,4}^{\mu\nu},\\
\end{gathered}}
which are the $(1,5)$ terms. The $(1,4)^\mu$ terms are
\eqn{\begin{gathered}
\sz_{8}^\mu=\cG_i^1\,\partial^\mu g^i\,E_4,\qquad
\sz_{9}^\mu=\cG_i^2\,\partial^\mu g^i\,I,\qquad
\sz_{10}^\mu=\tfrac{1}{25}\cG_i^3\,\partial^\mu g^i\,R^2,\\
\sz_{11}^\mu=\tfrac{1}{5}\cG_i^4\,\partial^\mu g^i\,\square R,\qquad
\sz_{12,\ldots,17}^\mu=\cH_i^{1,\ldots,6}\,\partial_\nu g^i
  H_{1,\ldots,6}^{\mu\nu}.
\end{gathered}}
The $(2,4)$ terms are given by
\eqn{\begin{gathered}
\st_{19}=\tfrac12\cG_{ij}^1\,\partial_\mu g^i\partial^\mu g^j\, E_4,\qquad
\st_{20}=\tfrac12\cG_{ij}^2\,\partial_\mu g^i\partial^\mu g^j\, I,\qquad
\st_{21}=\tfrac{1}{50}\cG_{ij}^3\,\partial_\mu g^i\partial^\mu g^j\,
  R^2,\\
\st_{22}=\tfrac{1}{10}\cG_{ij}^4\,\partial_\mu g^i\partial^\mu g^j\,
  \square R,\qquad
\st_{23,\ldots,28}=\tfrac12\cH_{ij}^{1,\ldots,6}\,\partial_\mu
  g^i\partial_\nu g^j\,H_{1,\ldots,6}^{\mu\nu},
\end{gathered}}
while the $(2,3)^\mu$ terms are
\eqn{\begin{gathered}
\sz_{18}^\mu=\cF_i\,\nabla_\kappa\partial_\lambda g^i\,\nabla^\mu
  G^{\kappa\lambda},\qquad
\sz_{19}^\mu=\tfrac15\cE_i\,\square g^i\,\partial^\mu R,\qquad
\sz_{20}^\mu=\tfrac15\cE_{ij}\,\partial^\mu g^i\partial_\nu
  g^j\,\partial^\nu R.
\end{gathered}}
The $(3,3)$ terms are
\eqn{\begin{gathered}
\st_{29}=\cF_{ij}\,\partial_\kappa g^i\nabla_\lambda\partial_\mu g^j\,
  \nabla^\kappa G^{\lambda\mu},\qquad
\st_{30}=\cF_{ij}^\prime\,\partial_\kappa g^i\nabla_\lambda\partial_\mu
  g^j\,\nabla^\lambda G^{\kappa\mu},\\
\st_{31}=\tfrac12\cF_{ijk}\,\partial_\kappa g^i\partial_\lambda
  g^j\partial_\mu g^k\,\nabla^\kappa G^{\lambda\mu},\qquad
\st_{32}=\tfrac15\hat{\cE}_{ij}\,\partial_\mu g^i\square g^j\,\partial^\mu
  R,\\
\st_{33}=\tfrac{1}{10}\cE_{ijk}\,\partial_\mu g^i\partial_\nu g^j
  \partial^\nu g^k\,\partial^\mu R,
\end{gathered}}
and the $(3,2)^\mu$ terms are
\eqn{\begin{gathered}
\sz_{21}^\mu=\cD_{ij}\,\partial_\kappa g^i\nabla_\lambda\partial_\nu g^j\,
  R^{\mu\lambda\kappa\nu},\qquad
\sz_{22}^\mu=\cC_i\,\partial_\nu\square g^i\,G^{\mu\nu},\qquad
\sz_{23}^\mu=\cC_{ij}\,\partial_\kappa g^i\nabla_\nu
  \partial^\kappa g^j\,G^{\mu\nu},\\
\sz_{24}^\mu=\cC_{ij}^\prime\,\partial_\nu g^i\square g^j\,G^{\mu\nu},
  \qquad
\sz_{25}^\mu=\tfrac15\cB_{ij}\,\partial^\mu g^i\square g^j\,R.
\end{gathered}}
The $(4,2)$ terms are
\eqn{\begin{gathered}
\st_{34}=\cD_{ijk}\,\partial_\kappa g^i \partial_\mu g^j \nabla_\lambda
  \partial_\nu g^k\,R^{\kappa\lambda\mu\nu},\qquad
\st_{35}=\tfrac14\cD_{ijkl}\,\partial_\kappa g^i \partial_\lambda g^j
  \partial_\mu g^k\partial_\nu g^l\,R^{\kappa\lambda\mu\nu},\\
\st_{36}=\hat{\cC}_{ij}\,\nabla_\mu\partial_\nu g^i\square
  g^j\,G^{\mu\nu},\qquad
\st_{37}=\tfrac12\hat{\cC}_{ij}^\prime\,\nabla_\kappa\partial_\mu g^i
  \nabla^\kappa\partial_\nu g^j\,G^{\mu\nu},\qquad
\st_{38}=\tfrac12\cC_{ijk}\,\partial_\mu g^i\partial_\nu g^j\square g^k\,
  G^{\mu\nu},\\
\st_{39}=\cC_{ijk}^\prime\,\partial_\mu g^i\partial_\kappa g^j\nabla^\kappa
  \partial_\nu g^k\,G^{\mu\nu},\qquad
\st_{40}=\tfrac12\cC_{ijk}^{\prime\prime}\,\partial_\kappa g^i
  \partial^\kappa g^j \nabla_\mu\partial_\nu g^k\, G^{\mu\nu},\\
\st_{41}=\tfrac14\cC_{ijkl}\,\partial_\mu g^i \partial_\nu g^j
  \partial_\kappa g^k \partial^\kappa g^l\,G^{\mu\nu},\qquad
\st_{42}=\tfrac15\cB_i\,\square^2 g^i\,R,\qquad
\st_{43}=\tfrac{1}{10}\hat{\cB}_{ij}\,\square g^i\square g^j\,R,\\
\st_{44}=\tfrac{1}{10}\hat{\cB}_{ij}^\prime\,\nabla_\mu\partial_\nu g^i
  \nabla^\mu\partial^\nu g^j\,R,\qquad
\st_{45}=\tfrac{1}{10}\cB_{ijk}\,\partial_\mu g^i\partial^\mu g^j \square
  g^k\,R,\\
\st_{46}=\tfrac{1}{10}\cB_{ijk}^\prime\,\partial_\mu g^i\partial_\nu g^j
  \nabla^\mu\partial^\nu g^k\,R,\qquad
\st_{47}=\tfrac{1}{20}\cB_{ijkl}\,\partial_\mu g^i\partial^\mu g^j
  \partial_\nu g^k\partial^\nu g^l\,R,
\end{gathered}}
and the $(5,0)^\mu$ terms are
\eqn{\begin{gathered}
\sz_{26}^\mu=\cA_{ij}\,\partial_\nu\square g^i \nabla^\mu\partial^\nu
  g^j,\qquad
\sz_{27}^\mu=\cA_{ij}^{\prime}\,\partial^\mu g^i\square^2 g^j,\qquad
\sz_{28}^\mu=\cA_{ijk}\,\partial_\nu g^i \nabla^\mu\partial^\nu
  g^j\square g^k,\\
\sz_{29}^\mu=\cA_{ijk}^{\prime}\,\partial_\kappa g^i
  \nabla^\mu\partial_\lambda g^j\nabla^\kappa\partial^\lambda g^k,\qquad
\sz_{30}^\mu=\tfrac12\cA_{ijkl}\,\partial_\nu g^i \partial^\nu
  g^j\partial^\mu g^k \square g^l.
\end{gathered}}
Finally, the $(6,0)$ terms are
\eqn{\begin{gathered}
\st_{48}=\cA_i\,\square^3 g^i,\qquad
\st_{49}=\hat{\cA}_{ij}\,\square^2 g^i\square g^j,\qquad
\st_{50}=\tfrac12\hat{\cA}_{ij}^\prime\,\partial_\mu\square g^i\partial^\mu
  \square g^j,\\
\st_{51}=\tfrac12\hat{\cA}_{ij}^{\prime\prime}\,\nabla_\kappa\nabla_\lambda
  \partial_\mu g^i\nabla^\kappa\nabla^\lambda\partial^\mu g^j,\qquad
\st_{52}=\tfrac18\hat{\cA}_{ijk}\,\square g^i \square g^j \square g^k,\\
\st_{53}=\tfrac12\hat{\cA}_{ijk}^{\prime}\,\nabla_\kappa\partial_\mu g^i
  \nabla^\kappa\partial_\nu g^j \nabla^\mu\partial^\nu g^k,\qquad
\st_{54}=\hat{\cA}_{ijk}^{\prime\prime}\,\partial_\mu g^i\square
  g^j\partial^\mu \square g^k,\\
\st_{55}=\check{\cA}_{ijk}\,\partial_\mu g^i\nabla^\mu\partial_\nu g^j
  \partial^\nu\square g^k,\qquad
\st_{56}=\tfrac12\check{\cA}_{ijk}^\prime\,\partial_\mu g^i\partial^\mu g^j
  \square^2 g^k,\\
\st_{57}=\tfrac12\check{\cA}_{ijk}^{\prime\prime}\,\partial_\mu g^i
  \partial_\nu g^j \nabla^\mu\partial^\nu\square g^k,\qquad
\st_{58}=\tfrac14\hat{\cA}_{ijkl}\,\partial_\mu g^i \partial^\mu g^j
  \square g^k\square g^l,\\
\st_{59}=\tfrac14\hat{\cA}_{ijkl}^\prime\,\partial_\kappa g^i
  \partial^\kappa g^j \nabla_\mu\partial_\nu g^k \nabla^\mu\partial^\nu
  g^l,\qquad
\st_{60}=\tfrac12\hat{\cA}_{ijkl}^{\prime\prime}\,\partial_\kappa g^i
  \partial_\lambda g^j \nabla^\kappa\partial_\mu g^k
  \nabla^\lambda\partial^\mu g^l,\\
\st_{61}=\tfrac12\check{\cA}_{ijkl}\,\partial_\mu g^i
  \partial_\nu g^j \nabla^\mu\partial^\nu g^k\square g^l,\qquad
\st_{62}=\tfrac12\check{\cA}^\prime_{ijkl}\,\partial_\kappa g^i
  \partial_\lambda g^j \partial_\mu
  g^k\nabla^\kappa\nabla^\lambda\partial^\mu g^l,\\
\st_{63}=\tfrac14\cA_{ijklm}\,\partial_\mu g^i \partial^\mu g^j
  \partial_\nu g^k \partial^\nu g^l \square g^m,\qquad
\st_{64}=\tfrac14\cA_{ijklm}^\prime\,\partial_\kappa g^i \partial^\kappa
  g^j \partial_\lambda g^k \partial_\mu g^l \nabla^\lambda\partial^\mu
  g^m,\\
\st_{65}=\tfrac18\cA_{ijklmn}\,\partial_\kappa g^i \partial^\kappa g^j
  \partial_\lambda g^k \partial^\lambda g^l \partial_\mu g^m \partial^\mu
  g^n.
\end{gathered}}

Note that when we have more than two derivatives on a coupling only one
ordering of the derivatives is independent. That is because all other
orderings can be produced by commuting covariant derivatives, a process
which introduces Riemann tensors or its contractions. This leads to terms
with curvature tensors that we have already included.

The scalar quantities $a, b_{1,\ldots,14}$ and $c_{1,2,3}$ are functions of
the couplings $g^i$. All $\cA,\ldots,\cI$ are also functions of the
couplings, but not all of them are tensors under reparametrizations in the
space of couplings, owing to the fact that $\square g^i$ transforms
inhomogeneously under $g^i\to\bar{g}^i(g)$; more specifically, $\square
\bar{g}^i=\partial_j\bar{g}^i\,\square g^j +\partial_j \partial_k
\bar{g}^i\,\partial_\mu g^j\partial^\mu g^k$.  $\cE_{ijk}$ in $\st_{33}$ is
an example of this, because of the $\square g^j$ in $\st_{32}$.

\bibliography{Wcc6d}

\begin{thebibliography}{10}
\ifx\href\asklfhas\newcommand{\href}[2]{#2}\fi
\ifx\arxivref\asklfhas\newcommand{\arxivref}[2]{\href{http://arxiv.org/abs/#1}{#2}}\fi
\ifx\doiref\asklfhas\newcommand{\doiref}[2]{\href{http://dx.doi.org/#1}{#2}}\fi
\parskip 0pt
\normalsize

\bibitem{Wess:1971yu}
J.~Wess \& B.~Zumino,
\textit{``{Consequences of anomalous Ward identities}''},
\doiref{10.1016/0370-2693(71)90582-X}{Phys.Lett. \textbf{B37}, 95 (1971)}.

\bibitem{Witten:1983tw}
E.~Witten,
\textit{``{Global Aspects of Current Algebra}''},
\doiref{10.1016/0550-3213(83)90063-9}{Nucl.Phys. \textbf{B223}, 422 (1983)}.

\bibitem{Osborn:1991gm}
H.~Osborn,
\textit{``{Weyl consistency conditions and a local renormalization group
  equation for general renormalizable field theories}''},
\doiref{10.1016/0550-3213(91)80030-P}{Nucl.Phys. \textbf{B363}, 486 (1991)}.

\bibitem{Zamolodchikov:1986gt}
A.~Zamolodchikov,
\textit{``{Irreversibility of the Flux of the Renormalization Group in a 2D
  Field Theory}''},
JETP~Lett. \textbf{43}, 730 (1986).

\bibitem{Jack:1990eb}
I.~Jack \& H.~Osborn,
\textit{``{Analogs for the c theorem for four-dimensional renormalizable field
  theories}''},
\doiref{10.1016/0550-3213(90)90584-Z}{Nucl.Phys. \textbf{B343}, 647 (1990)}.

\bibitem{Cardy:1988cwa}
J.L. Cardy,
\textit{``{Is There a c Theorem in Four-Dimensions?}''},
\doiref{10.1016/0370-2693(88)90054-8}{Phys.Lett. \textbf{B215}, 749 (1988)}.

\bibitem{Polchinski:1987dy}
J.~Polchinski,
\textit{``{Scale and conformal invariance in quantum field theory}''},
\doiref{10.1016/0550-3213(88)90179-4}{Nucl.Phys. \textbf{B303}, 226 (1988)}.

\bibitem{Fortin:2012hn}
J.F. Fortin, B.~Grinstein \& A.~Stergiou,
\textit{``{Limit Cycles and Conformal Invariance}''},
\doiref{10.1007/JHEP01(2013)184}{JHEP \textbf{1301}, 184 (2013)},
\normalsize{\texttt{\arxivref{1208.3674}{arXiv:1208.3674}}}.

\bibitem{Fortin:2012cq}
J.F. Fortin, B.~Grinstein \& A.~Stergiou,
\textit{``{Limit Cycles in Four Dimensions}''},
\doiref{10.1007/JHEP12(2012)112}{JHEP \textbf{1212}, 112 (2012)},
\normalsize{\texttt{\arxivref{1206.2921}{arXiv:1206.2921}}}.

\bibitem{Luty:2012ww}
M.A. Luty, J.~Polchinski \& R.~Rattazzi,
\textit{``{The $a$-theorem and the Asymptotics of 4D Quantum Field Theory}''},
\doiref{10.1007/JHEP01(2013)152}{JHEP \textbf{1301}, 152 (2013)},
\normalsize{\texttt{\arxivref{1204.5221}{arXiv:1204.5221}}}.

\bibitem{Fortin:2012hc}
J.F. Fortin, B.~Grinstein, C.W. Murphy \& A.~Stergiou,
\textit{``{On Limit Cycles in Supersymmetric Theories}''},
\doiref{10.1016/j.physletb.2012.12.059}{Phys.Lett. \textbf{B719}, 170 (2013)},
\normalsize{\texttt{\arxivref{1210.2718}{arXiv:1210.2718}}}.

\bibitem{Nakayama:2012nd}
Y.~Nakayama,
\textit{``{Supercurrent, Supervirial and Superimprovement}''},
\normalsize{\texttt{\arxivref{1208.4726}{arXiv:1208.4726}}}.

\bibitem{Fortin:2011ks}
J.F. Fortin, B.~Grinstein \& A.~Stergiou,
\textit{``{Scale without Conformal Invariance: An Example}''},
\doiref{10.1016/j.physletb.2011.08.060}{Phys.Lett. \textbf{B704}, 74 (2011)},
\normalsize{\texttt{\arxivref{1106.2540}{arXiv:1106.2540}}}.

\bibitem{Fortin:2011sz}
J.F. Fortin, B.~Grinstein \& A.~Stergiou,
\textit{``{Scale without Conformal Invariance: Theoretical Foundations}''},
\doiref{10.1007/JHEP07(2012)025}{JHEP \textbf{1207}, 025 (2012)},
\normalsize{\texttt{\arxivref{1107.3840}{arXiv:1107.3840}}}.

\bibitem{Fortin:2012ic}
J.F. Fortin, B.~Grinstein \& A.~Stergiou,
\textit{``{Scale without Conformal Invariance at Three Loops}''},
\doiref{10.1007/JHEP08(2012)085}{JHEP \textbf{1208}, 085 (2012)},
\normalsize{\texttt{\arxivref{1202.4757}{arXiv:1202.4757}}}.

\bibitem{Nakayama:2013wda}
Y.~Nakayama,
\textit{``{Consistency of local renormalization group in d=3}''},
\normalsize{\texttt{\arxivref{1307.8048}{arXiv:1307.8048}}}.

\bibitem{Barnes:2004jj}
E.~Barnes, K.A. Intriligator, B.~Wecht \& J.~Wright,
\textit{``{Evidence for the strongest version of the 4d a-theorem, via
  a-maximization along RG flows}''},
\doiref{10.1016/j.nuclphysb.2004.09.016}{Nucl.Phys. \textbf{B702}, 131 (2004)},
\normalsize{\texttt{\arxivref{hep-th/0408156}{hep-th/0408156}}}.

\bibitem{Komargodski:2011vj}
Z.~Komargodski \& A.~Schwimmer,
\textit{``{On Renormalization Group Flows in Four Dimensions}''},
\doiref{10.1007/JHEP12(2011)099}{JHEP \textbf{1112}, 099 (2011)},
\normalsize{\texttt{\arxivref{1107.3987}{arXiv:1107.3987}}}.

\bibitem{Elvang:2012st}
H.~Elvang, D.Z. Freedman, L.Y. Hung, M.~Kiermaier, R.C. Myers \& S.~Theisen,
\textit{``{On renormalization group flows and the a-theorem in 6d}''},
\doiref{10.1007/JHEP10(2012)011}{JHEP \textbf{1210}, 011 (2012)},
\normalsize{\texttt{\arxivref{1205.3994}{arXiv:1205.3994}}}.

\bibitem{Osborn:1991mk}
H.~Osborn,
\textit{``{Local renormalization group equations in quantum field theory}''},
in \textit{``Dubna 1991 Proceedings, Renormalization group '91''},
p.~128.

\bibitem{Capper:1974ic}
D.~Capper \& M.~Duff,
\textit{``{Trace anomalies in dimensional regularization}''},
\doiref{10.1007/BF02748300}{Nuovo~Cim. \textbf{A23}, 173 (1974)}.

\bibitem{Friedan:2009ik}
D.~Friedan \& A.~Konechny,
\textit{``{Gradient formula for the beta-function of 2d quantum field
  theory}''},
\doiref{10.1088/1751-8113/43/21/215401}{J.Phys. \textbf{A43}, 215401 (2010)},
\normalsize{\texttt{\arxivref{0910.3109}{arXiv:0910.3109}}}.

\bibitem{Bonora:1985cq}
L.~Bonora, P.~Pasti \& M.~Bregola,
\textit{``{Weyl Cocycles}''},
\doiref{10.1088/0264-9381/3/4/018}{Class.Quant.Grav. \textbf{3}, 635 (1986)}.

\bibitem{Bastianelli:2000rs}
F.~Bastianelli, G.~Cuoghi \& L.~Nocetti,
\textit{``{Consistency conditions and trace anomalies in six-dimensions}''},
\doiref{10.1088/0264-9381/18/5/303}{Class.Quant.Grav. \textbf{18}, 793 (2001)},
\normalsize{\texttt{\arxivref{hep-th/0007222}{hep-th/0007222}}}.

\bibitem{Lovelock:1971yv}
D.~Lovelock,
\textit{``{The Einstein tensor and its generalizations}''},
\doiref{10.1063/1.1665613}{J.Math.Phys. \textbf{12}, 498 (1971)}.

\bibitem{Deser1993279}
S.~Deser \& A.~Schwimmer,
\textit{``{Geometric classification of conformal anomalies in arbitrary
  dimensions}''},
\doiref{10.1016/0370-2693(93)90934-A}{Phys.Lett. \textbf{B309}, 279  (1993)},
\normalsize{\texttt{\arxivref{hep-th/9302047}{hep-th/9302047}}}.

\bibitem{Misner:1974qy}
C.W. Misner, K.~Thorne \& J.~Wheeler,
\textit{``{Gravitation}''},
{W.H.\ Freeman and Co.} (1973).

\end{thebibliography}

\end{document}